# Polarization splitter-rotator on thin film lithium niobate based on multimode interference

MENGKE WANG,[1] HAO YAO,[1] JIAYAO DENG,[1] ZHEFENG HU,[1] TINGTING TANG,[2] AND KAIXIN CHEN,[1]

[1]*School of optoelectronic Science and Engineering, University of Electronic Science and Technology of China, Chengdu 610054, China*
[2]*Optoelectronic Sensor Devices and Systems Key Laboratory of Sichuan Provincial University, Chengdu University of Information Technology, Chengdu, Sichuan, 610225, China*
*chenkx@uestc.edu.cn

**Abstract:** Polarization splitter-rotators (PSRs) are the key elements to realize on-chip polarization manipulation. Current PSRs on thin film lithium niobate (TFLN) rely on sub-micron gaps to realize modes separation, which increase the difficulties of lithography and etching. In this paper, a polarization splitter-rotator on TFLN based on multimode interference (MMI) is demonstrated. Mode division is achieved by an MMI-based mode demultiplexer. The feature size of the PSR is 1.5 μm, which can be fabricated with low priced i-line contact aligners. Experimental results show a polarization extinction ratio (PER) > 20 dB and insertion loss (IL) <1.5 dB are achieved in a wavelength range of 1542-1600 nm for TE-polarized light. And a PER > 9.5 dB and an IL <3.0 dB are achieved in a wavelength range of 1561-1600 nm for TM-polarized light. This PSR could find application in the low-cost fabrication of dual-polarization TFLN integrated photonic devices.

## 1. Introduction

Lithium niobate is a widely used optical material platform due to its wide transparency window, low optical loss, and outstanding linear electro-optical (EO) properties. In recent ten years, huge progresses have been made in integrated photonic devices based on thin film lithium niobate, such as frequency converters [1-3], optical frequency comb generators [4, 5], EO tunable interleaver [6], EO tunable mode (de)multiplexer [7], and EO modulators [8-12]. Among them, EO tunable devices and high-speed EO modulators receive most attention owing to their extensive application prospect. However, lithium niobate crystal has anisotropic EO coefficients, which makes TFLN EO devices highly polarization dependent. Thus, for TFLN based EO devices involving different polarization states, such as polarization independent EO modulators [13], dual-polarization in phase and quadrature (DP-IQ) modulators [11,12], and polarization management devices [14], on-chip polarization manipulation is a critical problem to be addressed.

   Polarization splitter-rotator (PSR) is the key element to realize on-chip polarization manipulation. Recently, several PSRs on TFLN platform have been demonstrated, which adopt same scheme based on mode conversion and division. Input light with arbitrary polarization state will excite $TE_0$ mode and $TM_0$ mode in the waveguide. Using an adiabatic taper, $TM_0$ mode could be converted to $TE_1$ mode, while $TE_0$ mode is unaffected. Then the $TE_0$ and $TE_1$ mode propagating in one waveguide need to be separated into two waveguides, which could be achieved by different kinds of mode demultiplexers, such as asymmetric directional coupler [15-18], asymmetric Y-junction [19], Y-junction combined with MMI [20] and adiabatic mode splitter [21]. However, all above devices contain narrow gaps of several hundred nanometers and thus need to be fabricated with EBL method [15-17, 19-21] or advanced steppers [18], which leads to low fabrication efficiency (EBL) and high equipment costs (EBL and stepper). As a contrast, mode demultiplexers based on multimode interference not only feature low loss, broad bandwidth and large fabrication tolerance [22-24], but also allow flexible design. By simply increasing the width of the multimode region, sub-micron structures can be avoided.

In this paper, we experimentally demonstrate an LNOI-based PSR utilizing cascaded two-stage MMIs to realize mode division. The feature size of the PSR is 1.5μm, which is much larger than other works [15-21] and can be easily fabricated with low priced i-line contact aligners. Moreover, the device shows relatively good performance. In a wavelength range of 1542 -1600 nm, for TE$_0$ polarized light, the insertion loss (IL) of the fabricated PSR is less than 1.5dB and the polarization extinction ratio (PER) is large than 20 dB. While for TM polarized light, the IL is less than 3.0 dB and the PER is larger than 9.5 dB over a wavelength range from 1561-1600 nm.

## 2. Structure and principles

Structure of the proposed PSR is shown in Fig.1(a), which consists of an adiabatic taper and a MMI-based mode demultiplexer. The cross-sectional view of the waveguide is shown in Fig.1(b). The thickness of the LN layer is $t_{LN}$ = 600 nm, the etch depth is $t$ = 200 nm and the sidewall angle $\theta$ is 70°. Fig.2 shows the effective indices of first three modes with different waveguide widths. It could be found that the mode hybridization region is around $w$ = 2.7 μm. To obtain larger fabrication tolerance, the start width $w_1$ and the end width $w_2$ of the adiabatic taper2 is set to be $w_1$ = 2.2 μm and $w_2$ = 3.2 μm. Width of the input waveguide is set to be $w_0$=2.0 μm and the length of taper1 is set to be $L_1$=50 μm. Width of the input port of MMI-1 is set to be $w_3$ = 3.5 μm and the length of taper3 is set to be $L_3$=100 μm. The calculated mode conversion efficiency (MCE) with various lengths of the adiabatic taper2 is given in Fig.3(a). It can be observed that MCE increases with the taper length $L_2$. To achieve complete mode conversion, $L_2$ is designed to be 800 μm. Simulated electric fields propagating in the taper2 are shown in Fig.3(b). For TE$_0$ mode input, the mode profile is not altered except for its size. While for TM$_0$ mode input, it could be clearly observed that TM$_0$ mode is converted to TE$_1$ mode.

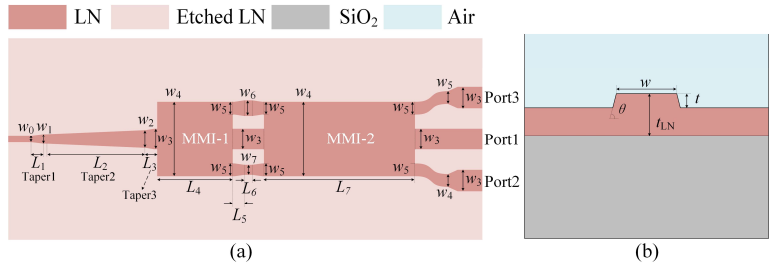

Fig. 1. (a) Schematic of the proposed PSR, (b) cross-sectional view of the LN waveguide

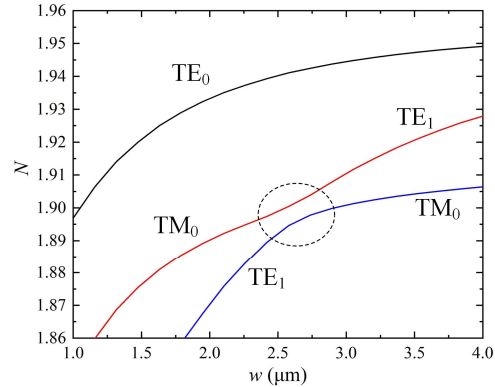

Fig. 2. Calculated dispersion curves of the LN waveguide as a function of the waveguide width at 1550 nm wavelength

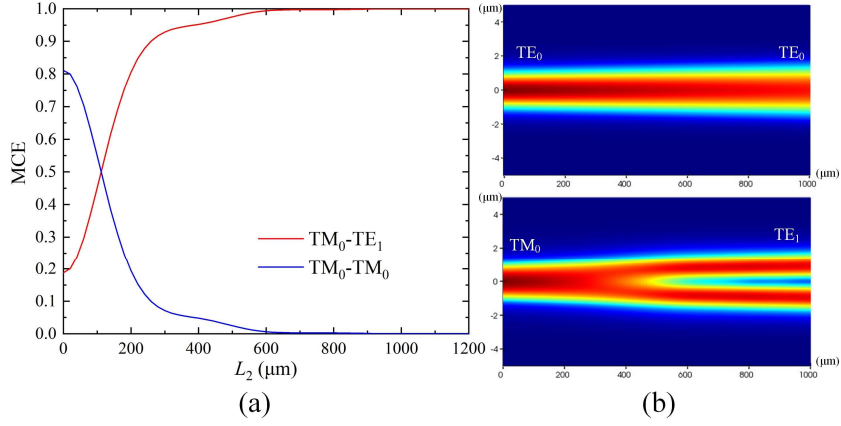

Fig. 3. (a) Calculated mode conversion efficiency as a function of the taper length $L_2$ at 1550 nm wavelength, (b)Simulated electricity filed profile of the taper2 for $TE_0$ input and $TM_0$ input at 1550 nm wavelength.

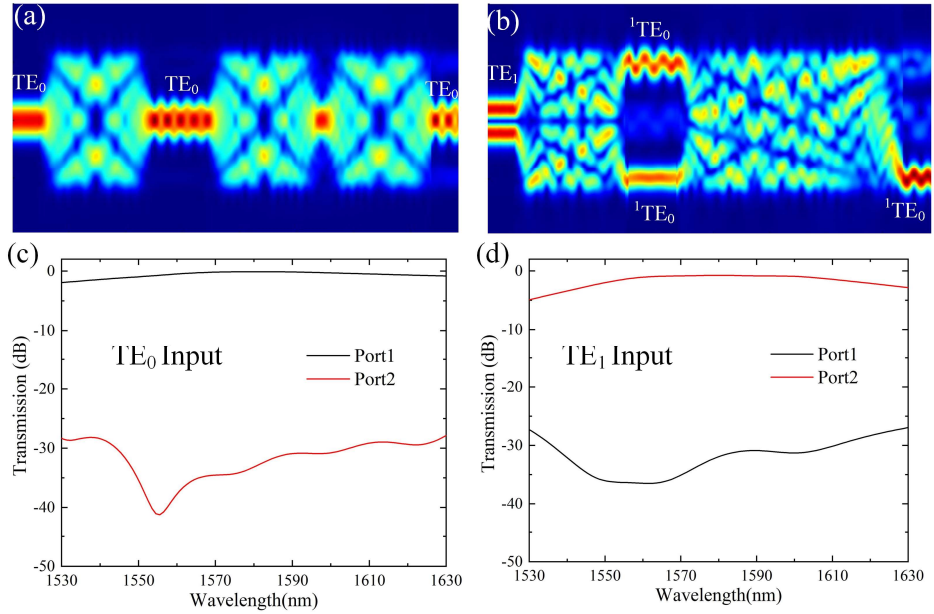

Fig. 4. Simulated electricity filed profile for (a) $TE_0$ input and (b)$TE_1$ input at 1550nm wavelength, calculated transmittance spectra of the MMI-based mode demultiplexer for (c) $TE_0$ input and (d) $TE_1$ input

The mode demultiplexer is composed of a 1×3 MMI (MMI-1), and a 3×3 MMI (MMI-2), as shown in Fig.1(a). Width of the two MMI are both set to be $w_4$ =11.8 μm. Lengths of MMI-1 and MMI-2 are designed to be $L_4$ = 198 μm and $L_7$ = 396 μm. Upper and lower output (input) ports of MMI-1 (MMI-2) have same width $w_5$ = 2.0 μm. Width of the upper/lower waveguide is set to be $w_6$ = 2.5 μm / $w_7$ =1.5 μm. Length of the taper connected to the upper/lower waveguide is set to be $L_5$ =25 μm. The simulated electric intensity profiles of the mode demultiplexer are shown in Fig.3(a). When $TE_0$ mode is launched into the mode demultiplexer, it will be mapped to the port1. When $TE_1$ mode is launched into the mode demultiplexer, it will be decomposed into two out-of-phase $^1TE_0$ modes at the upper and lower output ports of MMI-1. Since the upper waveguide and the lower waveguide have different width ($w_6$ = 2.5 μm and

$w_7$ = 1.5 μm), by changing their length $L_6$, the phase difference $\Delta\varphi$ of the two $^1TE_0$ modes could be tuned. Specially, for $\Delta\varphi=\pi/2$ (corresponding $L_6$ = 36 μm), two $^1TE_0$ modes will be recombined in MMI-2 and mapped to port2 [22, 23], as shown in Fig.3(b). The calculated transmittance spectra of the mode demultiplexer are shown in Fig.3(c) and Fig.3(d). For $TE_0$ mode input, over a wavelength range from 1550-1630nm, the insertion loss is lower than 1.0 dB and the extinction ratio is larger than 27 dB. While for $TM_0$ mode input, over a wavelength range from 1560-1610nm, the insertion loss is lower than 1.5 dB and the extinction ratio is larger than 30 dB.

### 3. Fabrication and measurement

The proposed PSR was fabricated on a commercial X-cut LNOI wafer (NANOLN) with 600nm thick LN layer. 500 nm thick photoresist (PR) was firstly coated on the wafer. The PR pattern was defined by an i-line contact aligner (SUSS MA-6). Then, the PR pattern was transferred to the LN layer by $Ar^+$ based inductively coupled plasma (ICP) etching process. After etching, the residual PR was removed and the chip facets was polished. A microscopic image of the MMI-based mode demultiplexer of the PSR is shown in Fig.5.

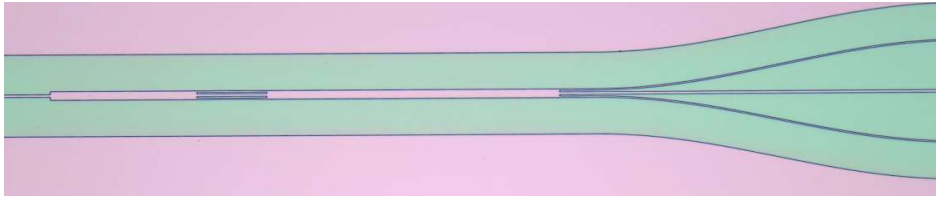

Fig. 5. Microscopic image of the MMI-based mode demultiplexer of the PSR.

Light from an amplified spontaneous emission (ASE) source (1530-1600nm) was polarized with an inline fiber polarizer. The polarization state of light was then adjusted by a polarization controller (PC) and coupled into the chip with a lensed single-mode fiber. The output light from the chip was collected with another single mode fiber and monitored with an optical spectrum analyzer (OSA) (YOKOGAWA, AQ6370D). The transmission spectra for $TE_0$ input are shown in Fig. 6(a). Over a wavelength range from 1542-1600nm, the IL is lower than 1.5 dB and the PER is larger than 20 dB. The transmission spectra for $TM_0$ input are shown in Fig. 6(a). Over a wavelength range from 1561-1600 nm, the IL is lower than 3.0 dB and the PER is larger than 9.5 dB.

Near-field images taken at 1550 nm by an infrared camera (MicronViewer 7290A) are shown in Fig.7. A linear polarizer is placed before the camera to capture the image of TE/TM polarized light. For $TE_0$ input, a strong $TE_0$ mode profile can be observed at port1, while almost no TM polarized light can be observed at three ports, which indicates a high PER. For $TM_0$ input, a strong $TE_0$ mode profile can be observed at port2, and a weak $TE_0$ mode profile can be observed at port3, which indicates the phase difference $\Delta\varphi$ of the two $^1TE_0$ modes has a small deviation from $\pi/2$. Additionally, weak TM polarized light can be observed at three ports and the slab region between them, which may be caused by the incomplete mode conversion in taper2. Thus, by further optimizing design parameters including $L_4$ (relate to $\Delta\varphi$), $L_2$, $w_1$ and $w_2$ (relate to MCE), IR can be reduced and PER can be improved for $TM_0$ input.

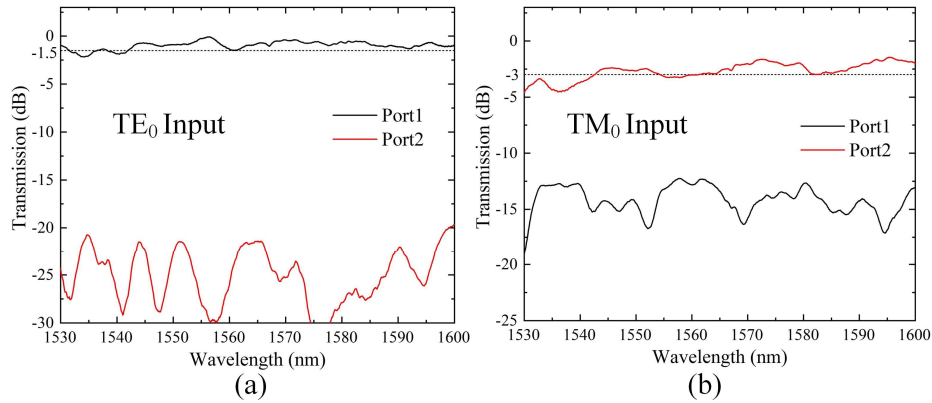

Fig. 6. Measured transmittance spectra of the PSR for (a) TE$_0$ input and (b) TM$_0$ input

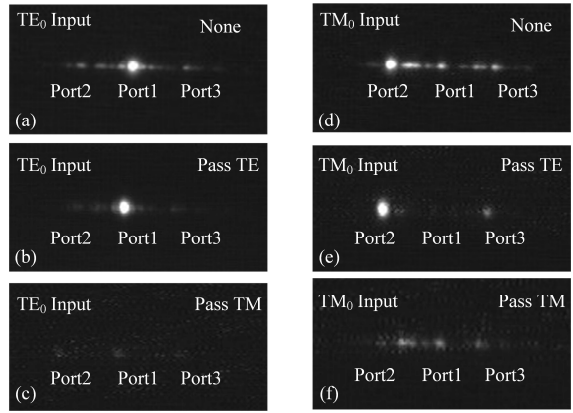

Fig. 7. Near field images taken at 1550 nm wavelength for TE$_0$ input with (a) no polarizer, (b) a polarizer passing TE polarized light, (c) a polarizer passing TM polarized light, and for TM$_0$ input with (d) no polarizer, (e) a polarizer passing TE polarized light, (f) a polarizer passing TM polarized light.

## 4. Conclusions

In this work, an LNOI polarization splitter and rotator based on multimode interference was demonstrated. The feature sized of our proposed device is 1.5 μm, which can be fabricated with low priced i-line contact aligner. Thus, the fabrication efficiency is significantly increased compared with PSRs fabricated with EBL method [15-17, 19-21] and the equipment costs are largely reduced compared with PSRs fabricated with i-line steppers [18]. Moreover, our fabricated PSR shows a low IL < 1.5 dB and high PER > 20 dB over a wavelength range from 1542-1600 nm for TE polarized light. Additionally, an IL <3.0 dB and PER > 9.5 dB over a wavelength range from 1561-1600 nm are achieved for TM polarized light, and the performances can be improved by further optimizing the design parameters. This PSR could find application in the low-cost fabrication of dual-polarization LNOI integrated photonic devices, such as DP-IQ LNOI modulators.

**Funding**

**Disclosures**

The authors declare no conflicts of interest.